\documentclass[12pt]{iopart}
\usepackage{amssymb}
\usepackage{theorem}
\usepackage{psfrag}

\usepackage[dvips]{graphicx}
\theoremstyle{plain}
\newcommand{\R}{\mathbb{R}}
\newcommand{\Z}{\mathbb{Z}}
\newcommand{\Fix}{\mathrm{Fix}\ }
\renewcommand{\vec}[1]{\mathbf{#1}}
\newcommand{\Hin}[1]{{H}^{\mathrm{in}}_{#1}}
\newcommand{\Hout}[1]{{H}^{\mathrm{out}}_{#1}}

\newcommand{\toinf}{\rightarrow\infty}

\renewcommand{\d}{\mathrm{d}}
\newcommand{\st}[1]{{{#1}^{\star}}}
\newcommand{\rar}{\rightarrow}
\newcommand{\Poincare}{Poincar\'{e} }
\newcommand{\eqref}[1]{(\ref{#1})}

\newcommand{\Hina}[1]{{H}^{\mathrm{in}}_{#1}}
\newcommand{\Houta}[1]{{H}^{\mathrm{out}}_{#1}}

\newcommand{\lpf}{\lambda_{\mathrm{pf}}}

\newcommand{\binom}[2]{{#1 \choose #2}}

\newcommand{\ignore}[1]{}

\newcommand{\xin}[2]{x^{\mathrm{in},#1}_{#2}}
\newcommand{\xout}[2]{x^{\mathrm{out,#1}}_{#2}}
\newcommand{\vin}[2]{v^{\mathrm{in},#1}_{#2}}
\newcommand{\vout}[2]{v^{\mathrm{out,#1}}_{#2}}

\newcommand{\ba}{\begin{eqnarray}}
\newcommand{\ea}{\end{eqnarray}}
\newcommand{\nn}{\nonumber}
\newtheorem{lemma}{Lemma}
\newtheorem{claim}{Claim}
\newtheorem{corollary}{Corollary}
\newtheorem{theorem}{Theorem}
\newtheorem{defn}{Definition}

\graphicspath{{../../figures/}{figures/}}

\begin{document}
\title{Resonance bifurcations from robust homoclinic cycles}
\author{Claire M Postlethwaite\dag\footnote[3]{Corresponding author (c.postlethwaite@math.auckland.ac.nz)}\ and Jonathan H P Dawes\ddag}

\address{\dag Department of Mathematics, University of Auckland, Private Bag 92019, Auckland 1142, New Zealand}
\address{\ddag Department of Mathematical Sciences, University of Bath, Claverton Down, Bath BA2 7AY, UK}

\begin{abstract}
We present two calculations for a class of robust homoclinic cycles with symmetry
$\Z_n\ltimes\Z_2^n$, for which the sufficient conditions for asymptotic
stability given by Krupa and Melbourne are not optimal.

Firstly, we compute optimal conditions for asymptotic stability using
transition matrix techniques which make explicit use of the geometry
of the group action.

Secondly, through an explicit computation of the global parts of the Poincar\'e map near the cycle
we show that, generically, the
resonance bifurcations from the cycles are supercritical:
a unique branch of asymptotically stable period orbits emerges from the
resonance bifurcation and exists for coefficient values where the cycle has
lost stability. This calculation
is the first to explicitly compute the criticality of a 
resonance bifurcation, and answers a conjecture of Field and Swift in a particular
limiting case. Moreover, we are able to obtain an
asymptotically-correct analytic expression for the
period of the bifurcating orbit, with no adjustable parameters, which
has not proved possible previously. We
show that the asymptotic analysis
compares very favourably with numerical results.
\end{abstract}


\submitto{Nonlinearity}

\section{Introduction}

Heteroclinic orbits between saddle-like invariant sets are of great interest
in dynamical systems since they generate many kinds of non-trivial
behaviour, including intermittency and chaotic dynamics \cite{K97}.
In continuous-time dynamical
systems (i.e.\ ordinary differential equations) a non-transversal
intersection may give rise to intermittent dynamics in the sense that
trajectories approach neighbourhoods of the saddle-like invariant sets
and spent a substantial amount of time there before moving rapidly to a
neighbourhood of the next invariant set.

It is well known that, while heteroclinic orbits in generic systems
of nonlinear ordinary differential equations (ODEs) are of codimension at
least one, in sets of ODEs containing invariant subspaces
they can exist for open sets of
parameter values, that is, they are codimension zero, and hence are referred
to as `robust'~\cite{K97,F96}. Three situations 
in which such invariant subspaces can arise are (i)
due to equivariance with respect to a symmetry group~\cite{KM95,KM02},
(ii) modelling assumptions such as the permanence of death in
Lotka--Volterra-type
models of population dynamics~\cite{ML75,H94,HS98}, and (iii)
structural restrictions such as
the coupled cell structures investigated recently by
Aguiar et al.~\cite{AADF09}.
Detailed results on the existence of homoclinic cycles in particular 
families of equivariant vector fields have been given by many
authors, for example~\cite{JP87,AGH88,GH88}.

A number of codimension-one bifurcations have been identified in which robust
heteroclinic cycles are created or destroyed, or in which their
stability changes. Issues of stability turn out to be more subtle than
might be at first thought. In particular, weaker definitions of stability than
asymptotic stability turn out to be useful, and indeed
for some systems describe the generic case. The most prevalent of these
weaker notions is `essential asymptotic stability', introduced by Melbourne~\cite{Mel91} 
who gave an example of an essentially asymptotically stable robust
heteroclinic cycle. Recent work by Driesse and Homburg~\cite{DH09a}
discusses examples
of essentially asymptotically stable homoclinic cycles which are
produced in codimension-one bifurcations from asymptotically
stable robust homoclinic cycles as a `transverse'
eigenvalue (which we shall define in section~\ref{sec:rhc})
crosses the imaginary axis.

In this paper we discuss a particularly simple class of homoclinic
cycles in $\R^n$.
Although necessarily restrictive, this class contains the essence of the
Lotka--Volterra type examples discussed at length by Rabinovich and
co-workers \cite{RVLHAL01,VVR05,RVSA06,RHVA08} as models for
neural decision-making processes.
The class of homoclinic cycles that we consider can undergo both transverse
and resonance bifurcations and we
focus here on the resonant case, beginning by proving necessary and sufficient
conditions for asymptotic stability, improving on the general
result given by Krupa and Melbourne~\cite{KM95,KM02}. Such a resonance
bifurcation
is shown to yield a single periodic orbit which lies close to the
(now unstable) homoclinic cycle. The stability and direction
of bifurcation of this periodic orbit depends on coefficients in
the Poincar\'e return map which come from the global maps, that is, those
which describe how trajectories near the homoclinic cycle behave outside
neighbourhoods of the equilibria on the cycle~\cite{Chos97}.
Usually these coefficients are impossible to compute
analytically, but in this case the calculation turns out to be
tractable, asymptotically in the limit where the sum of the
transverse eigenvalues is small. To the best of our knowledge no
calculation along these lines has been attempted previously.
The result of the calculation is to show that the
resonance bifurcation for this class of simple homoclinic cycles
is always supercritical, at least when the sum of the transverse
eigenvalues is sufficiently small compared to the leading
expanding and contracting eigenvalues.

Field and Swift~\cite{FS91} study in detail a particular example from
the class of cycles we consider, in the case $n=4$. They conjecture
that the resonant homoclinic bifurcation is always supercritical. Our
results prove this conjecture for an open set of parameter values.

The outline of the paper is as follows. In section~\ref{sec:rhc} we
introduce our notation and the class of robust homoclinic cycles that
we study. We state our result on asymptotic stability of such cycles
(Theorem 1). Section~\ref{sec:stab} contains the proof of Theorem 1.
In section~\ref{sec:crit} we present the calculation of the coefficients
in the Poincar\'e return map. The calculations are
reasonably straightforward to follow but become remarkably
lengthy. Some more detailed parts of the calculations are relegated to
the Appendix. In section~\ref{sec:numerics} we use our return map
calculations to provide an estimate of the period of the bifurcating
periodic orbit. We also show that the agreement between
the analytic leading-order
estimate and numerical results is very good. We conclude in
section~\ref{sec:disc}.

\section{Robust homoclinic cycles}
\label{sec:rhc}

We consider continuous time dynamical systems in the form of $\Gamma$-equivariant ODEs:
\begin{equation}
\label{eq:ode}
\dot{x}=f(x),\quad x\in\R^n
\end{equation}
where $\Gamma\subset\vec{O}(n)$ is a finite Lie group.
An equilibrium $\xi\in\R^n$ of~\eqref{eq:ode} satisfies $f(\xi)=0$.
We begin by giving a number of definitions;
these are all standard in the literature, see for example~\cite{K97,KM02}.

\begin{defn}\label{def:het_con}
$\phi_j(t)$ is a \emph{heteroclinic orbit} between two equilibria
$\xi_j$ and $\xi_{j+1}$ of~\eqref{eq:ode} if $\phi_j(t)$
is a solution of~\eqref{eq:ode} which is backward asymptotic to
$\xi_j$ and forward asymptotic to $\xi_{j+1}$.
\end{defn}

\begin{defn}
A \emph{heteroclinic cycle} is an invariant set $X\subset\R^n$ consisting of the
union of a set of equilibria $\{\xi_1,...,\xi_m\}$ and
orbits $\{\phi_1,...,\phi_m\}$, where $\phi_j$ is a
heteroclinic orbit between $\xi_j$ and $\xi_{j+1}$; and
$\xi_{m+1}\equiv\xi_1$. We require that $m\geq 2$. 
\end{defn}

In the case $m=1$, that is, $\xi_2=\xi_1$, we say that
$\phi_1$ is a \emph{homoclinic orbit} to $\xi_1$. 

\begin{defn}
A heteroclinic cycle is a \emph{homoclinic cycle} if there exists
$\gamma\in\Gamma$ such that 
$\gamma\xi_j=\xi_{j+1}$ for all $j$.
\end{defn}

\begin{defn}
For $x\in\R^n$ the \emph{isotropy subgroup} $\Sigma_x$ is 
\begin{equation}\label{eq:iso_sg} \Sigma_x=\{\sigma\in\Gamma:\sigma
  x=x\}.\end{equation} 
  \end{defn}
  
  \begin{defn}
For $\Sigma$ an isotropy subgroup of $\Lambda$, the
\emph{fixed-point subspace} $\Fix\Sigma$ is 
\begin{equation} \label{eq:fps} \Fix\Sigma=\{x\in\R^n:\sigma x=x\
  \forall \sigma\in\Sigma\}.\end{equation} 
  \end{defn}

\begin{defn}\label{def:het_cyc}
A heteroclinic cycle $X$ is \emph{robust} if for each $j$, $1\leq
j\leq m$, there exists
a fixed-point subspace, $P_j=\Fix\Sigma_j$ where
$\Sigma_j\subset\Lambda$ and
\begin{enumerate}
\item $\xi_{j}$ is a saddle and $\xi_{j+1}$ is a sink for the flow
restricted to $P_j$,
\item there is a heteroclinic connection from $\xi_{j}$ to $\xi_{j+1}$
  contained in $P_j$.
\end{enumerate}
\end{defn}

Importantly, robust heteroclinic cycles may occur as codimension-zero phenomena in
systems with symmetry. That is, they may exist for open sets of parameter values.
We define
$L_j\equiv P_{j-1}\cap P_j$ and clearly $\xi_j\in L_j$.
Following~\cite{KM95}, the eigenvalues of the linearisation of $f(x)$
  about each equilibrium can be classified according to the
subspaces in which the eigenspaces lie, as shown in
  table~\ref{tab:evals}.
\begin{table}
\caption{\label{tab:evals}Classification of eigenvalues. $P\ominus L$ denotes the
orthogonal complement in $P$ of the subspace $L$.}
\begin{indented}
\item[]\begin{tabular}{@{}ll}
\br
Eigenvalue class & Subspace \\  \mr Radial ($r$) &
$L_j\equiv P_{j-1}\cap P_j$ \\ Contracting ($c$) &
$V_j(c)=P_{j-1}\ominus L_j$ \\ Expanding ($e$) &
$V_j(e)=P_j\ominus L_j$
\\ Transverse ($s$) & $V_j(s)=(P_{j-1}+P_j)^{\perp}$ \\  
\br
\end{tabular}
\end{indented}
\end{table}

The specific differential equations we consider in this paper are of the form:
\begin{equation} \label{eq:ode1}\dot{x}=f(x)=\mu x +Q(x),\quad x\in\R^n \end{equation} 
where $\mu\in\R_+$ (and so, following a rescaling of time, can be set
  equal to 1),  $n\geq 3$, and $Q(x)$ is a 
  $\Gamma$-equivariant polynomial, that is,
\[
\gamma Q(x)=Q(\gamma x),\qquad \gamma\in\Gamma,\]
which contains only nonlinear terms.
The group $\Gamma$ is of the form
\begin{equation} \label{eq:group}\Gamma=\Z_n\ltimes\Delta_n,\end{equation}  
where $\Delta_n\equiv\Z_2^n$ acts on $\R^n$ as $n$ reflections $\kappa_j$, $j=1,\dots,n$: 
\begin{equation}\label{eq:kappa}
\kappa_j(x_1,\dots,x_j,\dots,x_n)=(x_1,\dots,-x_j,\dots,x_n),\end{equation}
The actions of the reflections $\kappa_j$ mean that every coordinate hyperplane
is a fixed-point subspace, and hence invariant under~\eqref{eq:ode1}.
The $\Z_n$ subgroup of $\Gamma$
is generated by a cyclic permutation element $\rho$ which acts as:
\begin{equation}\label{eq:rho}
\rho(x_1,\dots,x_n)=(x_n,x_1,\dots,x_{n-1}).\end{equation}
It is clearly sufficient to consider the dynamics restricted to the domain
\[\R^n_+=\{(x_1,\dots,x_n)\in\R^n|x_1,\dots,x_n\geq 0\}.\]

We now describe the robust homoclinic cycles $X$ that we study
in this paper. For now, we suppose that it is possible to construct a
vector field with the following properties. In section~\ref{sec:crit}
we give an explicit example of ODEs that contain such a cycle.
We make the following assumptions on the ODEs~\eqref{eq:ode1}:

\begin{itemize}
\item[{\bf (H1)}] There exist $n$ equilibria $\xi_j$, $j=1,\dots n$, related by the symmetry element $\rho$:
\[\xi_{j+1}=\rho\xi_j.\]
each with isotropy exactly $\Delta_{n-l}$, ($1\leq l \leq n-2$), and therefore with $l$ non-zero coordinates.
\item[{\bf (H2)}] The unstable manifold of $\xi_j$, $W^u(\xi_j)$, is one-dimensional, and has
isotropy $\Delta_{n-l-1}$, that is, points on the manifold have $(l+1)$ non-zero coordinates.
\end{itemize}

Without loss of generality, we can choose the non-zero coordinates of $\xi_1$ to be $x_1,\dots,x_l$, and the non-zero coordinates of $W^u(\xi_1)$ to be $x_1,\dots,x_{l+1}$. Therefore $\xi_1$ and $\xi_2$ are contained in the subspace $P_1$, spanned by $x_1,\dots,x_{l+1}$. We make the further assumption:

\begin{itemize}
\item[{\bf (H3)}] $\xi_2$ is the only sink in $P_1$ and the unstable manifold of $\xi_1$ is asymptotic to $\xi_2$.
\end{itemize}

Therefore the union of the $\xi_j$ and their unstable manifolds forms a homoclinic cycle we label $X$.

From the action of $\rho^{-1}$, it is clear that $P_n$ is spanned by $x_n,x_1,\dots,x_l$.
Using table~\ref{tab:evals} we deduce that
$V_1(c)$ is the $x_n$ axis and $V_1(e)$ is the $x_{l+1}$ axis. The radial
subspace of $\xi_1$ has $l$ dimensions, and so there are $n-l-2$ transverse directions. Since $\xi_j$ has a one-dimensional unstable manifold, all the transverse and radial eigenvalues are negative. 

We label the contracting eigenvalue (that is, the eigenvalue in the direction spanned by $V_j(c)$ at $\xi_j$) 
$-c<0$, and the expanding eigenvalue (the eigenvalue in the direction spanned by $V_j(e)$), $e>0$. We label the transverse eigenvalues $-s_{1},\dots,-s_{n-l-2}$ where the eigenvector corresponding to the eigenvalue
$-s_k$  at $\xi_j$ is in the $x_{l+j+k}$ direction, taking subscripts
modulo $n$ as appropriate. 

\begin{theorem}\label{th:stab}
The homoclinic cycle $X$ is asymptotically stable if 
\begin{equation}
\label{eq:res_cond}c+\sum_{j=1}^{n-l-2} s_{j}>e.\end{equation}
If $c+\sum_{j=1}^{n-l-2} s_{j}<e$ then the cycle is unstable.
\end{theorem}

When the stability of the cycle changes at $c+\sum_{j=1}^{n-l-2}
s_{j}=e$, the cycle undergoes a \emph{resonance bifurcation}. In this
bifurcation a unique long-period periodic orbit bifurcates; the
bifurcation may be supercritical (in which case the bifurcating
periodic orbit is stable and exists for parameters for which $X$ is
unstable) or subcritical (and the bifurcating periodic orbit is
unstable and exists for parameters for which $X$ is stable). It turns
out that the
criticality of the resonance bifurcation is determined by the
behaviour of trajectories
in the `global parts' of the flow near $X$ (i.e.\ the parts of the
flow away from neighbourhoods of the equilibria).
In the second half of this paper we give an example of a system which
contains a homoclinic cycle of the form described above, and we
explicitly calculate these constants. For the example we give, we show
that the bifurcation always occurs supercritically. The
calculation also allows us to compute the period of the bifurcating
stable periodic orbit.

\section{Proof of theorem~\ref{th:stab}}
\label{sec:stab}

This section is devoted to the proof of
theorem~\ref{th:stab}. The proof is divided into three sections; first
we construct return maps on Poincar\'{e} sections in the standard way
by dividing the flow into `local' and `global' parts. The return maps
describe the behaviour of trajectories asymptotically close to $X$. We then 
relate these maps to transition matrices, and finally 
deduce results on the eigenvalues of the transition matrices which
enable us to prove the theorem.

We remark that, although this result can be deduced from a theorem of
Hofbauer and Sigmund \cite{HS98} (see chapter 17, pages 220--232),
we see substantial value in the proof presented here since it is more 
transparent for the situation at hand, and it makes the
geometry of the invariant subspaces explicit.

\subsection{\Poincare maps}
\label{sec:Pmaps}

We define Poincar\'{e} intersections about $\xi_1$ in the
standard way: 
\[\Hina{1}:\{x_n=h^2\},\quad \Houta{1}:\{x_{l+1}=h^2\},
\]
for some small $h>0$. We consider an initial point on $\Hina{1}$:
\[
 \xin{1}{}=(\vin{1}{1},\dots,\vin{1}{l},\xin{1}{l+1},\dots,\xin{1}{n-1},h^2)
,\] 
where the $v_j$ are radial coordinates which are zero at $\xi_1$.
The `time of flight' for the trajectory to reach $\Houta{1}$ is 
\begin{equation}
T_1=-\frac{1}{e}\log\left(\frac{\xin{1}{l+1}}{h^2}\right). \label{eqn:time}
\end{equation}
Let the coordinates of the trajectory when it reaches $\Houta{1}$ be
\[\xout{1}{}=(\vout{1}{1},\dots,\vout{1}{l},\xout{1}{l+1},\dots,\xout{1}{n})
\]
then using the linearised flow at $\xi_j$, we have
\begin{eqnarray}
\vout{1}{j}=h^{-\frac{2r_j}{e}}\vin{1}{j}\left(\xin{1}{l+1}\right)^{\frac{r_j}{e}}, & j=1,\dots,l \nonumber \\
\xout{1}{l+1}= h^2 ,&& \label{eq:local} \\
\xout{1}{j} = h^{\frac{-2s_{j-l-1}}{e}}\xin{1}{j}\left(\xin{1}{l+1}\right)^{\frac{s_{j-l-1}}{e}}, & j=l+2,\dots,n-1 \nonumber \\
\xout{1}{n} = h^{2-\frac{2c}{e}}\left(\xin{1}{l+1}\right)^{\frac{c}{e}}. & \nonumber
\end{eqnarray}
This defines the local map near $\xi_1$.

We now compute the global part of the Poincar\'e map.
The isotypic decomposition of $\R^n$ with respect to the isotropy of
$W^u(\xi_1)$ is
\[\R^n=P_1\oplus V_1(s_{1})\oplus\dots\oplus V_1(s_{n-l-2})\oplus V_1(c) \]
where $V_1(s_{j})=\langle(0,\dots,0,x_{l+j+1},0,\dots,0)\rangle$ is the
eigenspace spanned by the eigenvalue $s_j$. Recall that $\dim
P_1=l+1$. The form of the global map 
from $\Houta{1}$ to a point $\xin{2}{}$ on $\Hina{2}:\{x_{1}=h^2\}$, up to linear terms, is therefore:
\begin{equation} \label{eq:global}
\fl
\left( \begin{array}{c} \xin{2}{1} \\ \vin{2}{2} \\ \vdots \\ \vin{2}{l+1} \\ \xin{2}{l+2}\\ \vdots \\ \xin{2}{n} \end{array} \right)
  =\left(\!\! \begin{array}{c} h^2 \\ \st{w_2} \\ \vdots \\ \st{w_{l+1}} \\ 0 \\ \vdots \\ 0 \end{array}\! \!\right)+
\left(\! \begin{array}{ccccccc} 0 & \cdots & 0   & 0 & 0 & \cdots & 0 \\
c_{21} & \cdots  & c_{2l} & c_{2l+1} & 0 &\cdots & 0 \\
\vdots &    & \vdots & \vdots & \vdots & & \vdots \\ 
c_{l+11} & \cdots & c_{l+1l}  & c_{l+1l+1} & 0 &\cdots & 0 \\ 
0 & \cdots &0&0&  c_{l+2l+2} & \cdots  & 0 \\
\vdots&&\vdots&\vdots&\vdots& \ddots & \vdots \\
0&\cdots&0&0&0&\cdots& c_{mm}  \end{array} \!\right)\!
\left(\! \begin{array}{c} \vout{1}{1} \\ \vdots \\ \vout{1}{l} \\ h^2 \\ \xout{1}{l+2}\\ \vdots \\ \xout{1}{n}
\end{array}\! \right) 
\end{equation}
where the $c_{ij}$ are $O(1)$ constants which depend on the global
flow. By the invariance of the coordinate planes, $c_{jj}>0$ for ${l+2}\leq j
\leq n$. In the section~\ref{sec:crit}, we find approximations for these $c_{jj}$ for a specific set of ODEs.
 The $\st{w_j}$ are $O(h^2)$ constants which define where the heteroclinic connection $W^u(\xi_1)$ hits the plane $\Hina{2}$. Note that $\xin{2}{1}=h^2$ by definition, and the radial coordinates on $\Hina{2}$ are $\vin{2}{2},\dots,\vin{2}{l+1}$.

We compose the local and global maps, and use the symmetry
$\rho^{-1}$ to map points on $\Hina{2}$ onto $\Hina{1}$, to give a return map $\phi$ on $\Hina{1}$.
That is, we associate the point $\rho^{-1}\xin{2}{}$ with a point on $\Hina{1}$, so 
\[
\phi(\xin{1}{})=\rho^{-1}(\xin{2}{}),
\]
or
\[
\phi \left( \begin{array}{c} \vin{1}{1} \\ \vdots \\ \vin{1}{l} \\ \xin{1}{l+1} \\ \xin{1}{l+2}\\ \vdots \\ \xin{1}{n} \end{array} \right)
=\rho^{-1} 
\left( \begin{array}{c}\xin{2}{1} \\ \vin{2}{2} \\ \vdots \\ \vin{2}{l+1} \\ \xin{2}{l+2}\\ \vdots \\ \xin{2}{n} \end{array} \right)
=\left( \begin{array}{c} \vin{2}{2} \\ \vdots \\ \vin{2}{l+1} \\ \xin{2}{l+2}\\ \vdots \\ \xin{2}{n} \\ \xin{2}{1} \end{array} \right).
\]

The radial ($v_j$) components are contracting, and do not affect the other components, so we only need consider 
an $n-l-1$ dimensional map $\psi$ that describes the dynamics of the flow near the homoclinic cycle:
\newcommand{\Ct}{{\tilde{C}}}
\begin{equation} \label{eq:psi}
\fl
 \psi \left(\!\! \begin{array}{c}  \xin{1}{l+1} \\ \xin{1}{l+2}\\ \vdots \\ \xin{1}{n-2} \\ \xin{1}{n-1} \end{array}\!\! \right) =
\left( \!\! \begin{array}{c} \xin{2}{l+2}\\ \xin{2}{l+3} \\ \vdots \\ \xin{2}{n-1} \\ \xin{2}{n}  \end{array} \!\! \right) = 
\left(\! \begin{array}{c}  \Ct_{l+2}\xout{1}{l+2} \\ \Ct_{l+3}\xout{1}{l+3}\\ \vdots \\ \Ct_{n-1}\xout{1}{n-1} \\ \Ct_n\xout{1}{n} \end{array} \! \right)
=
\left(\!\! \begin{array}{c}
 \Ct_{l+2}h^{-\frac{2s_{1}}{e}}\xin{1}{l+2}\left(\xin{1}{l+1}\right)^{\frac{s_{1}}{e}} \\
\Ct_{l+3}h^{-\frac{2s_{2}}{e}}\xin{1}{l+3}\left(\xin{1}{l+1}\right)^{\frac{s_{2}}{e}} \\
\vdots \\
\\
\Ct_{n-1}h^{-\frac{2s_{n-l-2}}{e}}\xin{1}{n-1}\left(\xin{1}{l+1}\right)^{\frac{s_{n-l-2}}{e}} \\
\Ct_n h^{2(1-\frac{c}{e})}\left(\xin{1}{l+1}\right)^{\frac{c}{e}} 
\end{array} \!\! \right)
\end{equation}
where $\Ct_j=c_{jj}$.
\newcommand{\nt}{p}
For convenience, we relabel the coordinates and constants. We write  $\nt=n-l-1$, 
\[y_j=x_{l+j},\ j=1,\dots,\nt,\] 
\[C_j=h^{-\frac{2s_j}{e}}\Ct_{l+j},\ \mathrm{for}\ 1\leq j\leq p-1,\quad C_{p}=h^{2(1-\frac{c}{e})}\Ct_{n},\]
and
\begin{equation}\label{eq:aj}
a_j=\frac{s_{j}}{e}, \ \mathrm{for}\ 1\leq j\leq p-1,\quad   a_{\nt}=\frac{c}{e},
\end{equation}
to reach:
\begin{equation}\label{eq:psi1}
\psi\left(\!\! \begin{array}{c} y_1 \\ \vdots \\ y_{\nt-1} \\ y_\nt \\
\end{array} \!\! \right) = 
\left(\!\! \begin{array}{c}
  C_1y_2y_1^{a_1} \\ \vdots 
\\ C_{\nt-1}y_{\nt}y_1^{a_{\nt-1}} \\
 C_\nt y_1^{a_\nt} \end{array} \!\!\right).
\end{equation}

The map $\psi$ has a fixed point at $y_1=\dots=y_\nt=0$ which corresponds to the homoclinic cycle in the flow. The map has a second fixed point, the $y_1$ coordinate of which satisfies
\[
y_1=\left(\prod_{j=1}^{\nt}C_j\right) \left(y_1\right)^{\delta},
\]
where \[\delta=\sum_{j=1}^{p} a_j=\frac{c+\sum_{j=1}^{p-1}s_j}{e}.\] Note that $\prod_{j=1}^{\nt}C_j = h^{2(1-\delta)}\prod_{j=l+1}^{n}\Ct_j$. This fixed point may correspond to a periodic orbit in the flow, and we discuss the existence and stability of this fixed point further in section~\ref{sec:crit} below.

In the next section we use transition matrix methods to determine the stability of the zero fixed point of $\psi$ and hence the stability of the homoclinic cycle in the flow.

\subsection{Transition matrices}
\label{sec:tmatrices}

Let $G$ be the set of mappings $g:\R^\nt\rightarrow\R^\nt$ that have at
lowest order the form
\[g(x_1,\dots,x_\nt)=(C_1x_1^{\alpha_{11}}x_2^{\alpha_{12}}\cdots
x_\nt^{\alpha_{1\nt}},\dots, C_\nt x_1^{\alpha_{\nt1}}\cdots x_\nt^{\alpha_{\nt\nt}})\]
for constants $\alpha_{ij}\geq 0$ and $C_i$ non-zero. $G$ is clearly closed under
composition. We define the \emph{transition matrix}~\cite{KM95,FS91}
of $g$ to be the $\nt\times \nt$ real
matrix $M(g)$ with entries $[M(g)]_{ij}=\alpha_{ij}$. It is easily
verified that if $g_1, g_2 \in G$, then
\[M(g_2\circ g_1)=M(g_2)M(g_1).\]
Any $g\in G$ has a fixed point at $x_1=\dots=x_\nt=0$.
The zero fixed point of the map $g$ will be stable if all the row sums of
$M(g)^N$ diverge to $+\infty$ as $N\toinf$. Conversely, if any of the
row sums of $M(g)^N$ tends to $0$, then the fixed point is unstable.

For the homoclinic cycles described above, the transition matrix
corresponding to the Poincar\'{e} map~\eqref{eq:psi1} is:
\[M(\psi)=A_\nt=\left( \begin{array}{ccccc} a_1 & 1 & 0 & \cdots & 0 \\
a_2 & 0 & 1 &  \cdots & 0 \\
\vdots & \vdots & \ddots & \ddots & \vdots \\
a_{\nt-1} & 0 & \cdots & 0  & 1 \\
a_\nt & 0 &  \cdots& 0 & 0 \end{array} \right),\]
with the $a_i$ as in~\eqref{eq:aj}.
We now recall the Perron--Frobenius
theorem. We write $M\cdot\vec{v}$ to be the product $M_{ij}v_j$ for a matrix $M$
with components $M_{ij}$ and vector $\vec{v}$ with components $v_j$. The row sums of $A_\nt^N$ can be written as $A_\nt^N\cdot\vec{1}$ where $\vec{1}$ is the vector $(1,1,\dots,1)^T$.

\begin{defn}
A real $\nt\times \nt$ matrix $M$ is \emph{primitive} if there exists an $N>0$ such that all
  entries of $M^N$ are strictly positive.
\end{defn}
By inspection, $A_\nt$ is primitive, since $a_i>0$ for all $i$, and so $(A_p)^p$ will have strictly positive entries.
\begin{theorem}[Perron--Frobenius]
If $M$ is a
  $\nt\times \nt$ non-negative primitive matrix, then there
  exists a unique and simple positive eigenvalue $\lpf$ which is dominant
  in the sense that $|\lambda|<\lpf$ for all other eigenvalues $\lambda$ of
  $M$. There exist right and left eigenvectors $\vec{u}, \vec{v}> \vec{0}$
  such that $M\cdot\vec{u}=\lpf\vec{u}$ and
  $\vec{v}^{\top}\cdot M=\lpf\vec{v}^{\top}$. If $\vec{u}$ and $\vec{v}$ are normalised
  such that $\vec{v}^{\top}\cdot\vec{u}=1$ then \[\lim_{N\rar\infty}(\lpf^{-N}M^N)=
  T\equiv \vec{u}\vec{v}^{\top}.\]
\end{theorem}

We now consider matrices of the form $A_\nt$, with $a_i>0$, and in the remainder of this section, 
prove the following lemma.

\begin{lemma}\label{lem:lpf}
 If $\sum_{j=1}^\nt a_j <1$, then the dominant eigenvalue $\lpf$ of $A_\nt$ satisfies $\lpf<1$. If $\sum_{j=1}^\nt a_j >1$ then $\lpf>1$.
\end{lemma}

Using the Perron--Frobenius theorem, we have 
\[\lim_{N\rar\infty} \lpf^{-N}A_\nt^N\cdot\vec{1}=
T\cdot\vec{1}=\vec{u^{\prime}}> \vec{0},\]
where $\vec{u^{\prime}}=\left(\sum_j v_j\right)
\vec{u}$.
If $\lpf<1$, then the row sums tend to zero, that is, $\lim_{N\rar\infty} A_\nt^N\cdot \vec{1}=0$, and so the zero fixed point of the map $\psi$ is unstable. If $\lpf>1$, then $A_\nt^N\cdot\vec{1}$
will thus be divergent to $+\infty$, and the zero fixed point of the map will be stable. Since the stability of the fixed point in the map corresponds to the stability of the homoclinic cycle in the flow, substituting for the $a_j$ from~\eqref{eq:aj} completes the
proof of theorem~\ref{th:stab}.

We now prove lemma~\ref{lem:lpf}.

\begin{claim}
The eigenvalues $\lambda$ of $A_\nt$ satisfy
\begin{equation}\label{eq:char_poly}
\lambda^\nt-\left(a_1\lambda^{\nt-1}+a_2\lambda^{\nt-2}+\dots
+a_{\nt-1}\lambda +a_\nt\right)=0.
\end{equation}
\end{claim}
{\bf Proof.} We will show by induction (on $\nt$) that 
\begin{equation} \label{eq:Dn1} \fl  D_\nt\equiv \det(A_\nt-\lambda
I)=(-1)^\nt\left[\lambda^\nt-\left(a_1\lambda^{\nt-1}+
a_2\lambda^{\nt-2}+\dots+a_{\nt-1}\lambda
+a_\nt\right)\right].\end{equation} 
Firstly, we have
\[D_1=a_1-\lambda=(-1)[\lambda-a_1].\]
Then
\[
D_k =\left| \begin{array}{ccccc} a_1-\lambda & 1 & 0 & \cdots & 0 \\
a_2 & -\lambda & 1 &  \cdots & 0 \\
\vdots & \vdots & \ddots & \ddots & \vdots \\
a_{k-1} & 0 & \cdots & -\lambda  & 1 \\
a_k & 0 &  \cdots& 0 & -\lambda \end{array}\right| =-\lambda D_{k-1}+(-1)^{k-1}a_k.
\]
Assuming~\eqref{eq:Dn1} for $\nt=k-1$ we have
\begin{eqnarray*}
\fl
D_{k+1}=-\lambda(-1)^k\left[\lambda^k-\left(a_1\lambda^{k-1}+
a_2\lambda^{k-2}+\dots+a_{k-1}\lambda +a_k\right)\right]+(-1)^ka_{k+1} \\
\lo=(-1)^{k+1}\left[\lambda^{k+1}-\left(a_1\lambda^k+
a_2\lambda^{k-1}+\dots+a_k\lambda +a_{k+1}\right)\right].
\end{eqnarray*}
Hence~\eqref{eq:Dn1} is satisfied for all $\nt\geq 1$. \hfill{$\Box$}

The roots of equation~\eqref{eq:char_poly} are the eigenvalues of
$A_\nt$. The Perron--Frobenius theorem states that there is at least
one positive root, the next claim relates the size of the positive root(s)
to the entries $a_j$ of $A_\nt$.

\begin{claim} \label{cl:lam}
Let $\lambda$ be a positive root of (\ref{eq:char_poly}). 
If $\sum_{j=1}^\nt a_j<1$ then $\lambda<1$; if $\sum_{j=1}^\nt a_j>1$
then $\lambda>1$.
\end{claim}
{\bf Proof.} Consider the case
$\sum_{j=1}^\nt a_j<1$, and suppose that $\lambda \geq 1$:
\[
 \lambda^\nt = a_1\lambda^{\nt-1}+a_2\lambda^{\nt-2}+\dots+a_{\nt-1}\lambda
+a_\nt\leq\lambda^{\nt-1}\sum_{j=1}^\nt a_j<\lambda^{\nt-1}\leq\lambda^\nt.
\]
This is a contradiction and hence
$\lambda<1$. Similarly, for the case $\sum_{j=1}^\nt a_j>1$, suppose that
$\lambda \leq 1$, and then
\[
 \lambda^\nt =a_1\lambda^{\nt-1}+a_2\lambda^{\nt-2}+\dots+a_{\nt-1}\lambda+a_\nt\geq\lambda^{\nt-1}\sum_{j=1}^\nt
a_j>\lambda^{\nt-1}\geq\lambda^\nt,
\]
again a contradiction, and hence $\lambda>1$.\hfill{$\Box$}

Claim~\ref{cl:lam} is true for all positive roots $\lambda$, therefore it is
certainly true for the dominant eigenvalue $\lpf$, completing the proof of lemma~\ref{lem:lpf}. \hfill$\Box$

\section{Bifurcation criticality computation}
\label{sec:crit}

In this section we give an example set of ODEs which contain a
homoclinic cycle of the type discussed in section~\ref{sec:rhc}.  For
these equations, we are able to explicitly compute the shape of the
heteroclinic connections between the equilibria, and use this to
calculate the unknown constants $\Ct_j$ in the global part of the
\Poincare map~\eqref{eq:psi} derived in section~\ref{sec:Pmaps}. This
allows us to compute the criticality of the resonance bifurcation and
in addition, the period of the resulting bifurcating orbit.

\subsection{System description}
\label{sec:desc}

The following equations satisfy conditions {\bf (H1)} -  {\bf (H3)} and are equivariant under the action of $\Gamma$ given in~\eqref{eq:group},~\eqref{eq:kappa} and~\eqref{eq:rho}.
\begin{equation}\label{eq:odes2} \fl
\dot{x}_j=x_j(1-\sum_{i=1}^nx_i^2 - c x_{j+1}^2-s_{n-3} x_{j+2}^2-\cdots-s_{1}x_{j-2}^2 +e x_{j-1}^2),\quad j=1,\dots,n
\end{equation}
for $x=(x_1,\dots,x_n)\in\R^n$, $c,e,s_j>0$, ($j=1,\dots,n-3$).
In addition, in this example, $l=1$, and $Q(x)$ is truncated at third order. We consider only $n\geq 4$. The case $n=3$ is the example of Guckenheimer and Holmes~\cite{GH88} and it is well known that the system truncated at third order has a degenerate resonance bifurcation at $c=e$.

The system~\eqref{eq:odes2} has $n$ equilibria on the coordinate axes we label $\xi_j$; the equilibrium $\xi_j$ has coordinates $x_j=1$, $x_k=0$, $k\neq j$. Each equilibrium is hyperbolic, and the  eigenvalues of the linearisation about $\xi_j$ are -2 (in the radial, that is, $x_j$ direction), $-c$ (in the $x_{j-1}$ direction), $e$ (in $x_{j+1}$ direction), and $-s_k$  in the $x_{k+j+1}$ direction ($k=1,\dots,n-3$). Note that a homoclinic cycle between the equilibria $\xi_j$ exists if $-c$ and $e$ are of opposite sign, but without loss of generality we consider only the case $c,e>0$. Thus $e$ is the expanding eigenvalue and $c$ is the contracting eigenvalue. The $-s_j$ are transverse eigenvalues; we assume $-s_j<0$.
By theorem~\ref{th:stab}, the homoclinic cycle is asymptotically stable if $c+\sum_{j=1}^{n-3}s_j>e$. Otherwise it is unstable. 

In section~\ref{sec:Pmaps} we constructed a \Poincare map~\eqref{eq:psi} on a section $\Hin{1}$, approximating the flow near the homoclinic cycle. This map has a non-trivial fixed point with $x_2$-coordinate given by
\begin{eqnarray}\label{eq:x2}
x_2=h^2\left(\prod_{j=l+2}^{n}\Ct_j\right)^{\frac{1}{1-\delta}}
\end{eqnarray}
where the $\Ct_j$ are defined in the map~\eqref{eq:psi}.

This point corresponds to a periodic orbit in the flow, branching from the resonant homoclinic bifurcation, only if it is small as $\delta\rar 1$. 
Let $C=\left(\prod_{j=l+2}^{n}\Ct_j\right)$. For $C<1$ the fixed point is small if $\delta<1$ (where the heteroclinic cycle is unstable), so the bifurcation is supercritical. Conversely for $C>1$ the orbit exists in $\delta>1$, and the  bifurcation is subcritical. In the following, we write $A=\log C$. Thus the bifurcation is supercritical if $A<0$.

\subsection{The solution for the heteroclinic trajectory}

We now compute an approximation for the form of the heteroclinic
trajectory near the resonance bifurcation, 
under the assumption that the difference between the
expanding and contracting eigenvalues, $e-c$, is small. We write
$e=c+\beta$ and take $|\beta|\ll 1$. We consider the heteroclinic
connection from $\xi_1$ to $\xi_2$ in the $x_1$-$x_2$ plane; all other
connections are symmetry-related. The equations in the
$x_1$-$x_{2}$ plane are:
\begin{eqnarray*}
\dot{x}_1&=x_1(1-(x_1^2+x_{2}^2)-c x_{2}^2), \\
\dot{x}_{2}&=x_{2}(1-(x_1^2+x_{2}^2)+ex_{1}^2). 
\end{eqnarray*}
For ease of computation, we make a change of variables, writing $u=x_1^2$, $v=x_{2}^2$:
\begin{eqnarray}
\dot{u}&=2u(1-(u+v)-c v)  \label{eq:xy1}  \\
\dot{v}&=2v(1-(u+v)+e u) \label{eq:xy2} 
\end{eqnarray}
The heteroclinic connection is a solution to~\eqref{eq:xy1}
and~\eqref{eq:xy2} with boundary conditions $\{u=0, v=1\}$ and
$\{v=0,u=1\}$. When $c=e$ ($\beta=0$), there exists an exact solution
$v=v_0(u)\equiv 1-u$.

Recall that at resonance,
\[c+s=e,\]
where $s=\sum_{j=1}^{n-3}s_j$. Since $e=c+\beta$, at resonance we also have
\[s=\beta.\]
Therefore, close to resonance, $s$ is $O(\beta)$, although it is
important to note that it is not necessary that each $s_j$ is
individually $O(\beta)$.
We will now look for approximate solutions to~\eqref{eq:xy1} and~\eqref{eq:xy2} when $\beta\neq 0$.
 Write 
 \begin{equation}\label{eq:vofu} v=v_0(u)+\beta v_1(u)+O(\beta^2).\end{equation} 
 Expanding equations~\eqref{eq:xy1} and~\eqref{eq:xy2} in powers of $\beta$, we find at $O(\beta)$:
\[
v_1'(u)-\frac{1}{cu(1-u)}v_1(u)=-\frac{1}{c},
\]
which gives
\begin{equation}\label{eq:y1}
v_1(u)=\frac{1}{c}\left(\frac{u}{1-u}\right)^{1/c}\int_u^1 \left(\frac{1-\xi}{\xi}\right)^{1/c} \d \xi.
\end{equation}
In the following we also find it convenient to write the heteroclinic
connection as a function $u(v)$, that is, we write the connection in
the form
\begin{equation}\label{eq:uofv}
u(v)=1-v+\beta u_1(v)+O(\beta^2).
\end{equation}
By rearranging~\eqref{eq:uofv} and substituting into~\eqref{eq:vofu} we find that $u_1(v)=v_1(1-v)$, that is
\begin{equation}\label{eq:u1}
u_1(v)=\frac{1}{c}\left(\frac{1-v}{v}\right)^{1/c}\int_0^v\left(\frac{\eta}{1-\eta}\right)^{1/c} \d \eta.
\end{equation}

\subsection{Constants in the global map}

We now use the approximated form of the homoclinic
connection~\eqref{eq:vofu} to compute the constants $\Ct_j$ in the
map~\eqref{eq:psi}. From~\eqref{eq:psi} we have
\begin{equation}\label{eq:Ct}
\Ct_j=\frac{\xin{2}{j}}{\xout{1}{j}},\quad j=3,\dots,n.
\end{equation}
Consider a trajectory close to the $x_1$-$x_2$ plane, so that all
other coordinates are small. Approximations to the
$x_3$-,$\dots$,-$x_n$ equations in~\eqref{eq:odes2} are therefore
(ignoring higher order terms in the small coordinates):
\begin{eqnarray}
\dot{x}_{3}&=x_{3}(1-(x_1^2+x_{2}^2)-s_1x_1^2+ex_{2}^2), \label{eq:ukp2} \\
 \dot{x}_{j}&=x_{j}(1-(x_1^2+x_{2}^2)-s_{j-2}x_1^2-s_{j-3}x_{2}^2),\quad j=4,\dots,n-1, \label{eq:ukpj} \\
 \dot{x}_{n}&=x_{n}(1-(x_1^2+x_{2}^2)-cx_1^2-s_{n-3}x_{2}^2). \label{eq:ukm1} 
\end{eqnarray}

We consider a trajectory which starts close to the heteroclinic cycle. This trajectory will approximately follow the heteroclinic trajectory in the $x_1$-$x_{2}$ plane, so we integrate along the trajectory, from $\Hout{1}$ to $\Hin{2}$. So from~\eqref{eq:ukp2} for instance, we find:
\begin{equation}
\int_{\Hout{1}}^{\Hin{2}}\frac{1}{x_{3}} \d x_{3}=\int_{\Hout{1}}^{\Hin{2}} (1-(x_1^2+x_{2}^2)-s_1x_1^2+e x_{2}^2)\ \d t. \label{eq:het_int1}
\end{equation}
We can compute the left hand side of~\eqref{eq:het_int1}:
\[
\int_{\Hout{1}}^{\Hin{2}}\frac{1}{x_{3}} \d x_{3}=\log \left(\xin{2}{3}\right)-\log\left(\xout{1}{3}\right)
=\log\left(\frac{\xin{2}{3}}{\xout{1}{3}}\right)=\log(\Ct_{3}),
\]
(where the last equality follows from~\eqref{eq:Ct}), and we find
similar expressions for the other $\log(\Ct_j)$ by integrating
equations~\eqref{eq:ukpj} and~\eqref{eq:ukm1}.

As noted at the end of section~\ref{sec:desc}, the criticality of the bifurcation
depends on the sign of $A$, where
\[
A=\log C=\log \left(\prod_{j=3}^{n}\Ct_j \right)=\sum_{j=3}^{n}\log\Ct_j.
\]
Summing the integrated forms of equations~\eqref{eq:ukp2} to~\eqref{eq:ukm1}  we find, again writing $x_1^2=u$, and $x_2^2=v$,
\[
\fl
A=\sum_{j=3}^n\log\Ct_j=\int_{\Hout{1}}^{\Hin{2}} \left((n-2)(1-(u+v))-cu-\sum_{j=1}^{n-3}s_j (u+v) +ev  \right) \d t
\]
where again the integral is taken along the heteroclinic connection in
the $x_1$-$x_2$ plane. Note that this expression does not depend
independently on the $s_j$; it only depends on the sum
$s=\sum_{j=1}^{n-3} s_j$.

We now rewrite the integral as an integral in $u$. Recall that the
heteroclinic connection can be written as $v(u)=1-u+\beta
v_1(u)+O(\beta^2)$, and $e=c+\beta$. From~\eqref{eq:xy1}, we have
\[\dot{u}=-2cu(1-u)-2\beta (c+1)u v_1(u)+O(\beta^2),\] so
\[
\frac{1}{\dot{u}}=\frac{-1}{2cu(1-u)}+\beta\frac{c+1}{2c^2}\frac{v_1}{u(1-u)^2}+O(\beta^2).
\]
To determine the bounds on the integral, note that on $\Hout{1}$,
$x_2^2=h^2$, so $v=h$, and the $u$-coordinate of the heteroclinic
connection is $u=1-h+\beta u_1(h)+O(\beta^2)$. On $\Hin{2}$, $u=h$.

Thus, expanding in powers of $\beta$ (recall that $s=O(\beta)$) we find:
\begin{eqnarray}
A&= \int_{\Hout{1}}^{\Hin{2}} \left((n-2)(1-(u+v(u)))-cu-s (u+v(u)) +ev  \right) \frac{1}{\dot{u}} \d u \nonumber \\ \nonumber
&=  \int_{1-h+\beta u_1(h)}^{h} \left\{-cu +c(1-u) -s+\beta\left[-(n-2-c)v_1(u)+(1-u) \right]\right\} \\ \nonumber
&\qquad\times \left\{-\frac{1}{2c u(1-u)}+\beta\left[\frac{c+1}{2c^2}\frac{v_1}{u(1-u)^2}\right]\right\} \d u+O(\beta^2), \\ \nonumber
\end{eqnarray}
which, after rearranging terms and tidying up, becomes
\begin{eqnarray}
\fl
2A  =  \int_{1-h+\beta u_1(h)}^{h}  \frac{u -(1-u)}{ u(1-u)}\ \d u -  \beta \int_{1-h}^h \frac{1}{cu}\ \d u + s \int_{1-h}^h \frac{1}{c u(1-u)}\ \d u \nonumber \\  \nonumber
 +\beta \int_{1-h}^h \frac{(n-2-c)v_1(u)}{cu(1-u)}+(1-2u)\frac{c+1}{c}\frac{v_1(u)}{u(1-u)^2}\ \d u +O(\beta^2), \\ \nonumber
\lo=  \int_{1-h+\beta u_1(h)}^{h} \frac{1}{1-u}-\frac{1}{u}\ \d u-  \frac{\beta}{c} \int_{1-h}^h \frac{1}{u}\ \d u + \frac{s}{c} \int_{1-h}^h \frac{1}{u}+\frac{1}{1-u}\ \d u \\ \nonumber
 +\frac{\beta}{c} \int_{1-h}^h \frac{(n-2-c)(1-u)+(1-2u)(c+1)}{u(1-u)^2}v_1(u)\ \d u +O(\beta^2), \\ \nonumber
 \lo= \log\left(1-\frac{\beta u_1(h)}{h}\right)+\log\left(1+\frac{\beta u_1(h)}{1-h}\right)+\frac{2s- \beta}{c}(\log h -\log(1-h))  \\  
 -\beta\frac{n-1}{c} \int_{h}^{1-h} \frac{v_1(u)}{u(1-u)}\d u+\beta\frac{c+1}{c} \int_{h}^{1-h}\frac{v_1(u)}{(1-u)^2}\ \d u +O(\beta^2). \label{eq:A}
\end{eqnarray}
Since the functions $u_1$ and $v_1$ are (in principal) known through
the integrals~\eqref{eq:u1} and~\eqref{eq:y1} the above expression
determines the sign of $A$, and hence yields the criticality of the resonance
bifurcation. The remainder of this section is devoted to the
computation of $A$ which is perhaps surprisingly algebraically complicated.

\subsection{Expansion of integrals}
\label{sec:exps}

There are two types of terms in $A$ which we cannot yet express explicitly: the terms involving $u_1(h)$ and the integrals involving $v_1(u)$.  We will compute both of these by calculating a power series expansion for $u_1$. The two integrals in $A$ which we need to compute are:
\begin{equation}\label{eq:in1}
\int_{h}^{1-h} \frac{v_1(u)}{u(1-u)}\d u\equiv \int_{h}^{1-h} \frac{u_1(w)}{w(1-w)} \d w,
\end{equation}
and
\begin{equation}\label{eq:in2}
\int_{h}^{1-h}\frac{v_1(u)}{(1-u)^2}\ \d u\equiv \int_{h}^{1-h}\frac{u_1(w)}{w^2}\ \d w.
\end{equation}
For convenience, we write $q=1/c$, then
\[
u_1(w)=q\left(\frac{1-w}{w}\right) ^q\int_0^w\left(\frac{\xi}{1-\xi}\right)^q \d\xi.
\]
In the integrals~\eqref{eq:in1} and~\eqref{eq:in2} we only need to know $u_1(w)$ for $w<1$, so we can write the factors $(1-\xi)^{-q}$ and $(1-w)^q$ as power series. In the following, $\binom{q}{k}$ is a generalised binomial coefficient, that is, 
\[
 \binom{q}{k}\equiv\frac{\Gamma(q+1)}{\Gamma(k+1)\Gamma(q-k+1)},
\]
where $\Gamma(q)$ is the usual Gamma function.
Substituting the expansions into $u_1(w)$ gives:
\begin{eqnarray}
 u_1(w)&=qw^{-q}\sum_{k=0}^\infty\binom{q}{k}(-w)^k\int_0^w\xi^q\sum_{k=0}^\infty\binom{-q}{k}(-\xi)^k \d\xi, \nonumber \\
&=qw^{-q}\left(\sum_{k=0}^\infty (-1)^k \binom{q}{k} w^k\right)\left(\sum_{k=0}^\infty(-1)^k\binom{-q}{k}
\int_0^w\xi^{q+k} \d\xi \right), \nonumber \\
&= qw^{-q}\left(\sum_{k=0}^\infty (-1)^k \binom{q}{k}w^k\right)\left(\sum_{k=0}^\infty(-1)^k\binom{-q}{k}
\frac{1}{q+k+1}w^{q+k+1}  \right), \nonumber \\
&=qw\sum_{n=0}^\infty c_nw^n \label{eq:x1_cn}
\end{eqnarray}
where
\begin{eqnarray}
 c_n&=\sum_{k=0}^n (-1)^{n-k} \binom{q}{n-k}(-1)^{k}\binom{-q}{k}
\frac{1}{q+k+1}, \nonumber \\
&=\sum_{k=0}^n \binom{q}{n-k}\binom{q+k-1}{k} \frac{(-1)^{n-k}}{q+k+1} \label{eq:cn_old}
\end{eqnarray}

\begin{lemma}\label{lem:cn}
The coefficient $c_n$ in the expression for $u_1(w)$ given in~\eqref{eq:x1_cn}, can be written as
 \begin{eqnarray}
  c_n&=-\frac{q}{q+1}\frac{1}{\binom{n+q+1}{n}}\equiv-\frac{q\Gamma(n+1)\Gamma(q+1)}{\Gamma(q+n+2)}, \label{eq:cn_new} \quad n\geq 1\\
 c_0&=\frac{1}{q+1}. \label{eq:c0}
 \end{eqnarray}
\end{lemma}
{\bf Proof.}
See appendix.

\bigskip

We now evaluate the integrals~\eqref{eq:in1} and~\eqref{eq:in2}. We first consider~\eqref{eq:in1}, and writing $u_1(w)=qw\sum_{k=1}^{\infty} c_k w^k$ we find:
\begin{eqnarray*}
 \int_h^{1-h} \frac{u_1(w)}{w^2} \d w & = q\int_h^{1-h} c_0 \frac{1}{w} +\sum_{k=1}^\infty c_k w^{k-1} \d w, \\
&=qc_0\left[\log w\right]_h^{1-h} +q\sum_{k=1}^\infty \left[\frac{c_k}{k}w^{k}\right]_h^{1-h}, \\
&=qc_0(-\log h) +q\sum_{k=1}^\infty\frac{c_k}{k} +O(h).
\end{eqnarray*}
In the appendix we show that
\[
 \sum_{k=1}^\infty\frac{c_k}{k} = -\frac{q}{(q+1)^2},
\]
 and so, using this result together with~\eqref{eq:c0}, we have
\begin{equation}\label{eq:int1}
 \int_h^{1-h} \frac{u_1(w)}{w^2} \d w = -\frac{q}{q+1}\log h -\frac{q^2}{(q+1)^2} +O(h).
\end{equation}
Now we consider~\eqref{eq:in2}. We expand both $u_1(w)$ and $(1-w)^{-1}$ to find:
\begin{eqnarray*}
  \int_h^{1-h} \frac{u_1(w)}{w(1-w)} \d w& = q\int_h^{1-h} \left( \sum_{k=0}^\infty w^k\right)\left( \sum_{k=0}^\infty c_k w^k \right) \d w, \\
& = q\int_h^{1-h} \sum_{k=0}^\infty d_k w^k \d w, \\
& = q\left[ \sum_{k=0}^\infty  \frac{d_k}{k+1} w^{k+1} \right]_{h}^{1-h}, \\
& = q \sum_{k=0}^\infty \frac{d_k}{k+1} +O(h),
\end{eqnarray*}
where we have defined $d_k =\sum_{j=0}^k c_j$. We evaluate $d_k$ in the appendix to find:
\begin{eqnarray*}
\sum_{k=0}^\infty \frac{d_k}{k+1}& = \frac{1}{q}.
\end{eqnarray*}
Hence
\begin{equation}\label{eq:int2}
   \int_h^{1-h} \frac{u_1(w)}{w(1-w)} \d w = 1 +O(h).
\end{equation}

Finally, we use our expansion of $u_1$ to find $u_1(h)=qh\sum_{k=0}^\infty c_k h^k$, and so
\begin{equation}\label{eq:log1}
 \log\left(1-\frac{\beta u_1(h)}{h}\right)=-\beta q c_0 +O(\beta h) = -\beta \frac{q}{1+q}  +O(\beta h)
\end{equation}
and
\begin{equation}\label{eq:log2}
 \log\left(1-\frac{\beta u_1(h)}{1-h}\right)=O(\beta h).
\end{equation}

Therefore, substituting~\eqref{eq:int1},~\eqref{eq:int2},~\eqref{eq:log1} and~\eqref{eq:log2} into~\eqref{eq:A} we have
\begin{eqnarray*}
\fl 2A  = -\beta \frac{q}{q+1} +q(2s-\beta)\log h -\beta q (n-1) 
 +\beta (q+1) \left(\frac{q}{q+1}(-\log h)-\frac{q^2}{(q+1)^2}\right) \\ +O(\beta h) +O(\beta^2), \\
\lo = 2q(s-\beta)\log h+\beta\left(-\frac{q}{q+1} -nq +q
-\frac{q^2}{q+1} \right) +O(\beta h) +O(\beta^2), 
\end{eqnarray*}
which simplifies dramatically to give
\begin{eqnarray*}
A & = q(s-\beta)\log h -\frac{\beta nq}{2} +O(\beta h) +O(\beta^2).
\end{eqnarray*}
By definition, as we approach the bifurcation point $s\rightarrow
\beta$, and so, evaluating $A$ at the bifurcation point we obtain to leading order
\begin{eqnarray}\label{eq:a_expr}
 A\rightarrow -\frac{\beta n q}{2} <0,
\end{eqnarray}
and hence the bifurcation is found to be supercritical for all values of
$n \geq 4$ and $q=1/c$. Note
that although our calculation only computes $A$ to leading order in
$\beta$, since the criticality only depends on the sign of $A$, the
bifurcation will be supercritical whenever $\beta$ is sufficiently small.

\section{Bifurcating periodic orbits} 
\label{sec:numerics}

In this section we discuss the evolution of the unique stable periodic orbit
produced in the resonance
bifurcation for the system considered in section~\ref{sec:crit}. 

First we use the calculation of the
return map to derive an approximate expression for the period of the orbit.
We emphasise that the analytic result contains no adjustable parameters.
Referring back to the calculation of the return map,
combining equations~(\ref{eqn:time}) for the time spent in a neighbourhood of
an equilibrium on the cycle with~(\ref{eq:x2}) which gives the leading-order relation
between the location of the fixed point of the Poincar\'e map and the coefficients $C_j$, we obtain
the following expression for the period $P$ of the periodic orbit, as usual neglecting the time spent
moving between neighbourhoods of the equilibria on the cycle:
\ba
P \equiv n T = -\frac{n}{e} \log \left( C ^{1/(1-\delta)} \right), \nn
\ea
where $\delta=(c+s)/e$ and $C=\prod_{j=l+2}^n \tilde C_j$ as
before. Substituting $\log C \equiv A = -n \beta q/2$ where
$\beta=e-c>0$ and $q=1/c$ this expression simplifies to yield
\begin{equation}
P  =  \frac{n}{e} \frac{n \beta q}{2(1-\delta)} = \frac{n^2(e-c)}{2c(e-c-s)}. \label{eqn:p}
\end{equation}
This expression has been derived near the resonance bifurcation (i.e. $|e-c-s| \ll 1$), in the case
that $|e-c|\ll 1$ and therefore is expected to be asymptotically correct in the limits of small $s$
and small $e-c$.

We compare the analytic
result~(\ref{eqn:p}) with numerical integrations of~(\ref{eq:odes2}). Our numerical integrations
are carried out with a standard 4th order Runge--Kutta scheme, with the return times
$P_1,P_2,\ldots$ to the cross-section $x_1=0.2$ computed by linearly
interpolating points on the trajectory that lie on either side
of the cross-section. Since trajectories may be assumed to
converge exponentially to the periodic orbit (discounting
the possibility that it is nonhyperbolic) a highly accurate
extrapolation of the true period
of the orbit may be obtained by employing the \textit{ansatz}
$P_k=P + b_0\mathrm{e}^{-b_1 k}$ and eliminating the coefficients
(which are supposed to be constant) $b_0$ and $b_1$ by considering a set of three
return times $P_k$, $P_{k+m}$ and $P_{k+2m}$. This leads to the extrapolation formula
\begin{equation}
P  =  \frac{P_{k+2m} P_k - P_{k+m}^2}{P_{k+2m}-2P_{k+m}+P_k}, \label{eqn:extrap}
\end{equation}
which we find to give excellent results, even when the computed return times $P_j$ used
in~(\ref{eqn:extrap}) are far from the true (ultimate) period $P$.

\begin{figure}
\psfrag{P}{$P$}
\psfrag{e-c-s}{\hspace{-0.5cm}$e-c-s$}
\begin{center}
\includegraphics[width=11.0cm]{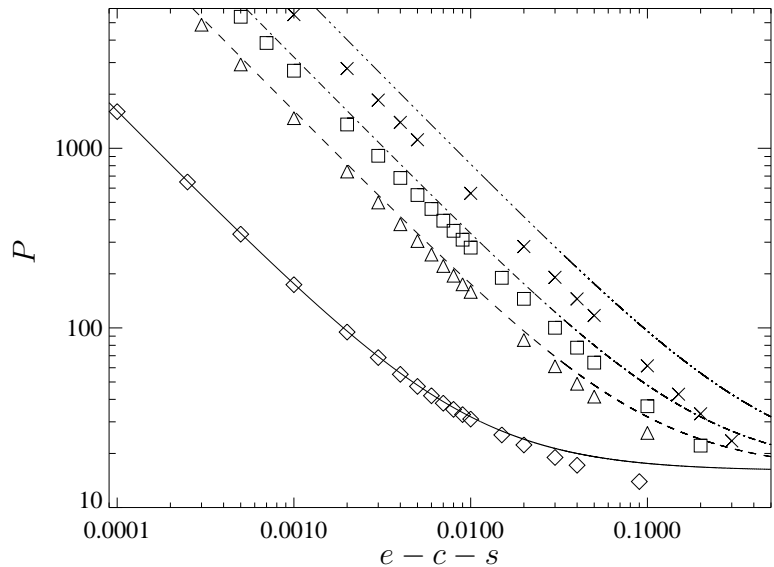}
\caption{\label{fig:varys}
Orbit period $P$ plotted against the eigenvalue combination $e-c-s$ for
the periodic orbit created in the resonance bifurcation at $e-c-s=0$.
Parameters are $n=4$, $c=0.5$. Calculations are carried out by varying
$e$ for fixed values of $s$.
Numerical data and the approximation~(\ref{eqn:p}) are shown for
four values of $s$, from bottom to top: $s=0.01$ ($\Diamond$, solid
line); $s=0.1$ ($\triangle$, dashed line);
$s=0.2$ ($\Box$, dash-dotted line); $s=0.5$ ($\times$,
dash-triple-dotted line). The 
approximation is excellent for small $s$ but systematically
overestimates the period for larger $s$.}
\end{center}
\end{figure}

\begin{figure}
\psfrag{P}{$P$}
\psfrag{e-c-s}{\hspace{-0.5cm}$e-c-s$}
\begin{center}
\includegraphics[width=11.0cm]{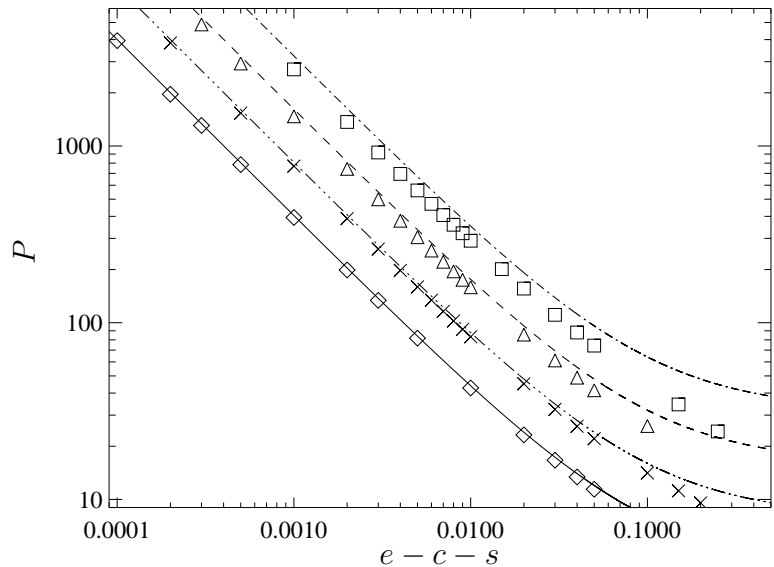}
\caption{\label{fig:varyc}
Orbit period $P$ plotted against the eigenvalue combination $e-c-s$ for the periodic orbit
created in the resonance bifurcation at $e-c-s=0$. Calculations are
carried out by varying $e$ for fixed values of $c$. 
Parameters are $n=4$, $s=0.1$. Numerical data and the approximation~(\ref{eqn:p}) are shown for
four values of $c$, from bottom to top: $c=2$ ($\Diamond$, solid line); $c=1$
($\times$, dash-triple-dotted line);
$c=0.5$ ($\triangle$, dashed line); $c=0.25$ ($\Box$, dash-dotted line). The
approximation is excellent for large $c$ but systematically overestimates the period for smaller $c$. Note
that the data for $c=0.5$ ($\triangle$, dashed line) is shown with the same symbols and
line style in figures~\ref{fig:varys} and~\ref{fig:varyn}.}
\end{center}
\end{figure}

\begin{figure}
\psfrag{P}{$P$}
\psfrag{e-c-s}{\hspace{-0.5cm}$e-c-s$}
\begin{center}
\includegraphics[width=11.0cm]{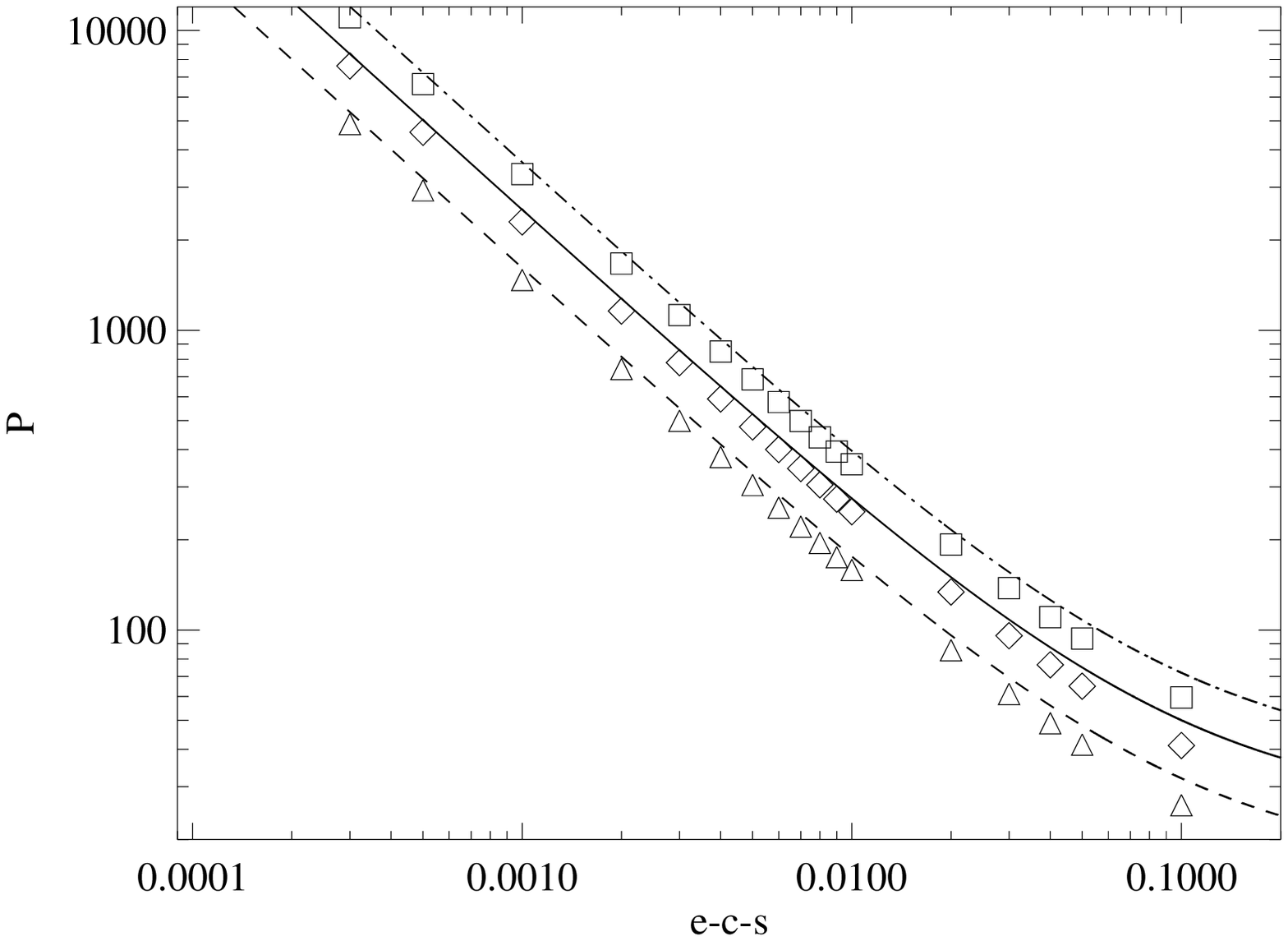}
\caption{\label{fig:varyn}
Orbit period $P$ plotted against the eigenvalue combination $e-c-s$
for the periodic orbit
created in the resonance bifurcation at $e-c-s=0$.
Calculations are carried out by varying
$e$ for fixed values of $s=0.1$ and $c=0.5$, keeping all the $s_j$
equal as discussed in the text.
Numerical data and the approximation~(\ref{eqn:p}) are shown for
three values of $n$, from bottom to top: $n=4$ ($\triangle$, dashed
line); $n=5$ ($\Diamond$, solid line);
$n=6$ ($\Box$, dash-dotted line). No trend in increasing or decreasing
accuracy is observed as $n$ varies. Note
that the data for $n=4$ ($\triangle$, dashed line) is shown with the same symbols and
line style in figures~\ref{fig:varys} and~\ref{fig:varyc}.}
\end{center}
\end{figure}

Figure~\ref{fig:varys} confirms the accuracy of the approximation~(\ref{eqn:p}),
showing the period $P$ as a function of $e-c-s$ for four
different values of $s$ in the case $n=4$ (where there is only a single transverse eigenvalue). The
excellent agreement
for $s=0.01$ (the solid curve and diamond symbols) and $s=0.1$ (dashed curve and triangle symbols) shows
that when $0<s \ll c$ the approximate expression~(\ref{eqn:p}) is extremely accurate.
Figure~\ref{fig:varyc} confirms the accuracy in the limit $0 < s \ll c$ when $c$ is varied at fixed $s$;
the data for $c=1$ are most accurately approximated by~(\ref{eqn:p}). Finally, figure~\ref{fig:varyn}
illustrates the dependence of $P$ on $n$ as being $P \sim n^2$ - not a scaling that one might intuitively
have proposed.
For this figure we have for convenience taken $s=0.1$ for $n=4,5,6$ and achieved this by
setting $s_j=s/(n-3)$ for all $1 \leq j \leq n-3$.
We remark that the accuracy of~(\ref{eqn:p}) does not appear to vary significantly with either increasing
or decreasing $n$.

Further numerical investigations in the case that $s>0$ but allowing one, or more, of the $s_j$ to be
negative, so that the transverse eigenvalues, $-s_j$ are positive, (i.e.\  when $e-c-s<0$, the homoclinic cycle is only essentially asymptotically stable rather
than asymptotically stable) show that a periodic orbit bifurcates at $e-c-s=0$ and has a period
that is extremely close to that for the $s_j=s/(n-3)$ case as long as $s$ remains much less than $e$.
For example, in the case $n=5$, $s=s_1+s_2=0.1$, $c=0.5$, $e=0.7$ we observed that $P=41.169$ for
$s_1=s_2=0.05$. On increasing $s_1$ (and correspondingly decreasing $s_2$ so that $s_1+s_2=0.1$ always)
we observed numerically that the period of the periodic orbit remained within $P=41.17 \pm 0.04$
for $0.05 \leq s_1 \leq 0.65$ before reducing rapidly for larger $s_1$: $P=35.4$ at $s_1=0.655$
and $P=30.15$ at $s_1=0.66$. It seems reasonable to expect that the shape of the periodic orbit
near the equilibria changes substantially when transverse eigenvalues become of
the same order as the expanding eigenvalue $e$. 
We leave a detailed investigation of the dynamics to be the subject of
future work.

\section{Discussion\label{sec:disc}}

In this paper we have discussed resonance bifurcations in a class of
vector fields containing robust homoclinic cycles that connect
equilibria on a single group orbit, each lying in an
$l$-dimensional hyperplane in $\R^n$ .
Using the well-known return map technique, and the construction
of transition matrices, we proved a general stability result for this
class of homoclinic cycles in section~\ref{sec:stab}.
We then discussed the resonant bifurcation
that occurs when $c+s=e$, where $s$ is the sum of the relevant
transverse eigenvalues.

In the case that the equilibria lie on the axes in $\R^n$ it is
possible to make substantial further progress and compute the period
of the bifurcating orbit by integrating along trajectories to
explicitly compute the global parts of the return map. To the best of
our knowledge, this is the first calculation of its kind.

The calculation rests on the assumption that the sum $s$ of the relevant
transverse eigenvalues is small compared to the leading stable and
unstable eigenvalues ($c$ and $e$ respectively). The details of the
calculation become algebraically rather complicated, but only involve
computing a number of integrals; this is achieved through power series
expansions and summations. Unlike most return map calculations where
the global parts of the map introduce undetermined coefficients whose
magnitudes are rarely known, our results contain no free parameters
and are seen to agree well with numerical computations of the period
of the bifurcating orbit.

It is of interest that the calculation shows that for this class of
systems the resonance bifurcation is always supercritical. It would be
of interest to investigate extending the present work to find a
correspondingly simple class of examples where the criticality of the
bifurcation depended non-trivially on parameters. For instance, in~\cite{PD06} we give an example of a cycle which has a resonant bifurcation, the criticality of which depends on parameters. It would also be
interesting to see to what extent the analysis here could be applied
to heteroclinic cycles in Lotka--Volterra systems, such as those
studied, mainly numerically, by other authors \cite{RVLHAL01,VVR05,RVSA06,RHVA08}.

\appendix

\section*{Appendix}
\setcounter{section}{1}

In this appendix we provide the proof of lemma~\ref{lem:cn}, and the computation of the sums $\sum_{k=1}^\infty \frac{c_k}{k}$ and $\sum_{k=0}^{\infty} \frac{d_k}{k+1}$ which are used in section~\ref{sec:exps} in the computation of $A$.

\paragraph{Proof of lemma~\ref{lem:cn}}

It is trivial to check that $c_0=1/(q+1)$. For $n\geq 1$, let
 \[
 f(n)=\sum_{k=0}^n F(n,k)
 \]
 where
 \[
 F(n,k)=\frac{1+q}{q} \frac{(-1)^{n-k+1}}{q+k+1} \binom{q+k-1}{k}\binom{q}{n-k}\binom{n+q+1}{n},
 \]
that is, $f(n)$ is the right hand side of~\eqref{eq:cn_old} divided by the right hand side of~\eqref{eq:cn_new}. We will show that $f(n)$ is independent of $n$, and specifically, that $f(n)=1$, which proves the lemma.

Note that $F(n,k)=0$ for $k\geq n+1$. Further, let
 \[
 R(n,k)=\frac{-k(1+k+q)(k-n+q)(k-1-n-nq)}{qn(1+n)^2(1-k+n)}
 \]
 for $k\neq n+1$, and define $G(n,k)=R(n,k)F(n,k)$. Note that $G(n,n+1)$ is well defined, and $G(n,k)=0$ for $k\geq n+2$.
 It can be verified that
 \[
 F(n+1,k)-F(n,k)=G(n,k+1)-G(n,k)
 \]
 Summing over all $k$ gives
 \begin{eqnarray}
\fl \sum_{k=0}^{\infty}  F(n+1,k)-\sum_{k=0}^{\infty}F(n,k)&=\sum_{k=0}^{\infty}G(n,k+1)-\sum_{k=0}^{\infty}G(n,k) \\
\fl \sum_{k=0}^{n+1}  F(n+1,k)-\sum_{k=0}^{n}F(n,k)&=\sum_{k=0}^{n}G(n,k+1)-\sum_{k=1}^{n+1}G(n,k) -G(n,0) \\
 f(n+1)-f(n)&=-G(n,0)=0
 \end{eqnarray}
 so $f(n)$ is independent of $n$. It is simple to check that $f(1)=1$, completing the proof. \hfill$\Box$

\medskip

The following lemma is used a number of times in what follows:
\begin{lemma}\label{lem:lem3}
\begin{equation}\label{eq:lem3}
 \sum_{j=0}^k\frac{\Gamma(x+1)\Gamma(j+1)}{\Gamma(x+j+1)} = \frac{x}{x-1}\left(1-\frac{\Gamma(x)\Gamma(k+2)}{\Gamma(x+k+1)}\right)
\end{equation}
\end{lemma}
{\bf Proof.}
We use induction on $k$. It is simple to check that~\eqref{eq:lem3} holds for $k=1$. Then assuming~\eqref{eq:lem3} is true for some $k>1$, we find for $k+1$,
\begin{eqnarray*}
\fl \sum_{j=0}^{k+1}\frac{\Gamma(x+1)\Gamma(j+1)}{\Gamma(x+j+1)} &= \sum_{j=0}^k\frac{\Gamma(x+1)\Gamma(j+1)}{\Gamma(x+j+1)} +\frac{\Gamma(x+1)\Gamma(k+2)}{\Gamma(x+k+2)} \\
&=\frac{x}{x-1}\left(1-\frac{\Gamma(x)\Gamma(k+2)}{\Gamma(x+k+1)}\right)+\frac{x\Gamma(x)\Gamma(k+2)}
{(x+k+1)\Gamma(x+k+1)} \\
&= \frac{x}{x-1}\left(1-\frac{\Gamma(x)\Gamma(k+2)}{\Gamma(x+k+1)}\left(\frac{k+2}{x+k+1} \right) \right) \\
& = \frac{x}{x-1}\left(1-\frac{\Gamma(x)\Gamma(k+3)}{\Gamma(x+k+2)}\right)
\end{eqnarray*}
\hfill$\Box$
\begin{corollary} \label{cor:cor} For $x>1$, 
\[
 \sum_{j=0}^\infty\frac{\Gamma(x+1)\Gamma(j+1)}{\Gamma(x+j+1)} =\frac{x}{x-1}
\]
\end{corollary}
{\bf Proof.} 
The required result is equivalent to showing that
\begin{eqnarray*}
\lim_{k \rar \infty} \frac{\Gamma(x)\Gamma(k+3)}{\Gamma(x+k+2)} &= 0.
\end{eqnarray*}
To show this we write the combination of Gamma functions as a Beta
function:
\begin{eqnarray}\label{eq:beta1}
\frac{\Gamma(x)\Gamma(k+3)}{\Gamma(x+k+2)} = (k+2)
B(k+2,x) = (x-1)B(k+3,x-1),
\end{eqnarray}
where we define the usual Beta function
\begin{eqnarray}\label{eq:beta}
B(p,q) &= \int_0^1 t^{p-1} (1-t)^{q-1} \, dt = \int_0^\infty
\frac{s^{q-1}}{(1+s)^{p+q}} \, ds,
\end{eqnarray}
making the change of variable $1+s=t^{-1}$, see \cite{WW27}. The last equality
in~\eqref{eq:beta1} can be derived in a straightforward manner by
integrating by parts the first integral 
expression in~\eqref{eq:beta} for $B(p,q)$. Then we have
\begin{eqnarray*}
\fl
(x-1) B(k+3,x-1) &= \int_0^\infty \frac{(x-1)s^{x-2}}{(1+s)^{x+k+2}}
\, ds < \int_0^\infty (x-1) (1+s)^{-k-4} \, ds, \\
&< -\left. \frac{(x-1)(1+s)^{-k-3}}{k+3}\right|_{s=0}^\infty = \frac{x-1}{k+3}, \\
\end{eqnarray*}
which clearly tends to zero as $k \rar \infty$ for any fixed $x>1$.
\hfill$\Box$

We are now able to compute the following two sums which are used in section~\ref{sec:exps}.
Firstly, consider
\begin{eqnarray*}
\fl
 \sum_{k=1}^\infty\frac{c_k}{k} = -q\sum_{k=1}^\infty\frac{\Gamma(k+1)\Gamma(q+1)}{k\Gamma(q+k+2)}, \\
\lo = -\frac{q}{(q+2)(q+1)}
\sum_{k=1}^\infty\frac{\Gamma(k)\Gamma(q+3)}{\Gamma(q+k+2)} 
= -\frac{q}{(q+2)(q+1)}  \sum_{k=0}^\infty\frac{\Gamma(k+1)\Gamma(q+3)}{\Gamma(q+k+3)} ,\\
 \lo = -\frac{q}{(q+2)(q+1)} \frac{q+2}{q+1}
= -\frac{q}{(q+1)^2}.
\end{eqnarray*}
Now consider
\begin{eqnarray*}
 d_k & =\sum_{j=0}^k c_j ,\\
& = c_0-q \sum_{j=1}^k\frac{\Gamma(j+1)\Gamma(q+1)}{\Gamma(q+j+2)}, \\
& = \frac{1}{1+q}+\frac{q\Gamma(q+1)}{\Gamma(q+2)} -q\sum_{j=0}^k\frac{\Gamma(j+1)\Gamma(q+1)}{\Gamma(q+j+2)}, \\
& = \frac{1}{1+q}+\frac{q}{1+q} -\frac{q}{q+1}\sum_{j=0}^k\frac{\Gamma(j+1)\Gamma(q+2)}{\Gamma(q+j+2)} ,\\
& = 1-\frac{q}{(q+1)}\frac{(q+1)}{q}\left(1-\frac{\Gamma(q+1)\Gamma(k+2)}{\Gamma(q+k+2)}\right) ,\\
& = \frac{\Gamma(q+1)\Gamma(k+2)}{\Gamma(q+k+2)},
\end{eqnarray*}
so that the second sum can be computed to be
\begin{eqnarray*}
\sum_{k=0}^\infty \frac{d_k}{k+1}& = \sum_{k=0}^\infty \frac{\Gamma(q+1)\Gamma(k+1)}{\Gamma(q+k+2)} ,\\
& = \frac{1}{q+1} \sum_{k=0}^\infty \frac{\Gamma(q+2)\Gamma(k+1)}{\Gamma(q+k+2)}, \\
& = \frac{1}{q+1} \frac{q+1}{q}
= \frac{1}{q}.
\end{eqnarray*}

\section*{Acknowledgements}

CMP acknowledges support from the University of Auckland Research
Council. JHPD currently holds a University Research Fellowship from
the Royal Society.
Preliminary work this topic was partially funded by the Nuffield
Foundation through an Undergraduate Research Bursary for Miss Yin Lin
(grant number 2006/33001) and by Newnham College, Cambridge.

\end{document}